# Achieving Long Retention in Area-Dependent Resistive Memory with Phase-Separated Amorphous Tantalum Oxide

*Sangyong Lee[1], Anton V. Ievlev[2], Dongjae Shin[1], Jingxian Li[1], and Yiyang Li[1]**

[1]Materials Science and Engineering, University of Michigan, Ann Arbor, MI 48109, USA

[2]Center for Nanophase Materials Science, Oak Ridge National Laboratory, Oak Ridge, TN, USA

*Corresponding authors: Yiyang Li; yiyangli@umich.edu

**Contact information:**

Sangyong Lee: sangylee@umich.edu

Anton V. Ievlev: ievlevav@ornl.gov

Dongjae Shin: dongjaes@umich.edu

Jingxian Li: jxli@umich.edu

Yiyang Li: yiyangli@umich.edu

## *Summary*

Resistive random-access memory (ReRAM) is a promising future nonvolatile memory technology. Most ReRAM exhibit a fundamental tradeoff: filament-type ReRAM provides long data retention but suffers from poor uniformity and high switching current, whereas nonfilamentary ReRAM shows lower-current, area-dependent switching but generally poor retention. No two-terminal device has been able to overcome this tradeoff. In this work, we present a two-terminal $Ta_2O_5$/ $TaO_X$ resistive memory cell that achieves both nonfilamentary switching and long retention. Electrical measurements and composition depth profile show that the switching is not confined to a single filament but is instead uniform across the entire switching region. Despite the nonfilamentary switching, this device can retain information for over 22 hours at 190 °C, which is comparable to the best filamentary devices. We propose that this long retention arises from composition phase separation in amorphous tantalum oxide. Our work shows that the fundamental tradeoff between information retention and switching uniformity can be overcome, and thereby provides a pathway toward more uniform and reliable oxide memory devices.

## *Introduction*

Resistive random-access memories (ReRAM) are promising next-generation non-volatile memories due to their simple metal-insulator-metal (MIM) and crossbar structure[1], fast switching speed[2], and potential for neuromorphic computing applications[1,3–7]. Among various material systems, valence change memory (VCM)[8] based on transition metal oxides is the most developed due to reliable switching and compatibility with CMOS processes[9]. Three types of valence-change memory have been proposed: filament, interface, and electrochemical[10]. While each has its advantages and disadvantages, none can meet all the reliability demands of advanced memory cells.

Filamentary VCM based on tantalum and hafnium oxides is the most widely studied. They switch through the formation and rupture of localized conductive filaments resulting from the movement of oxygen anions (or equivalently, oxygen vacancies)[11–13]. Because conduction is dominated by a single nanosized filament[13,14], the resistance of the device is independent of the geometric device size. Filamentary devices exhibit fast switching[15] and long retention greater than 10 years[8,13,15–19] and have been integrated into large crossbar arrays[20–23]. However, the stochastic nature of filament formation and rupture yields non-uniform switching behavior[16], device-to-device variability, and high operation current, usually greater than 100 micro-amps for switching[24]. The high operational current, especially for the RESET process, not only expends additional energy but also requires the use of larger transistors with a high enough driving current to control the current in a one-transistor-one-memristor (1T1R) geometry[25].

Nonfilamentary devices are an alternative VCM that solves the challenges of filamentary devices; examples of switching materials include $Pr_{1-x}Ca_xMnO_3$(PCMO) [26–32], $TiO_2$[33], $WO_3$[34–36], and $Ta_2O_5$[37–39]. Rather than forming localized filaments, such devices switch through oxygen exchange reactions, Schottky barrier modulation[40–43], charge trapping/de-trapping at the electrode–oxide interface[44,45], and bulk compositional change[46–48]. Because oxygen migration occurs uniformly across an entire area rather than a nanosized filament, the electronic conductance is proportional to the device area. This feature not only enables lower-conductance devices by reducing the device area but also improves device switching uniformity. However, such devices suffer from poor retention, typically on the order of a few minutes to hours[16,34–39], as the programmed states lose information over time. This poor retention is believed to be a result of the rapid migration of oxygen vacancies away from the interface. The ability to identify a device that retains nonfilamentary, area-dependent switching and excellent retention time has been a grand challenge in the field of resistive memory, especially one that can be achieved using back-end-of-line, CMOS-compatible materials.

In this study, we develop a two-terminal $Ta_2O_5$-based switching memory that is simultaneously nonfilamentary and nonvolatile. By increasing the thickness of the $Ta_2O_5$ switching layer to 15 nm, filament formation is suppressed. Our devices show both an electronic conductance that depends on the device area and a change in the bulk composition of the device under different resistance states, both of which are consistent with nonfilamentary switching. Despite the absence of filaments, information retention exceeds 24 hours at 190°C, a value similar to that of filamentary memory; this retention time projects to ten years at 85°C. We propose that this information retention results from composition-phase separation in tantalum oxide, which suppresses oxygen redistribution in the absence of an electrical bias; this mechanism has previously been shown to enable nonvolatile information retention in both filamentary ReRAM[13] and three-terminal electrochemical memory using tantalum oxide[49–51], tungsten oxide[52], and vanadium oxide[53]. Our results definitively demonstrate that it is possible to create a two-terminal resistive memory device that is simultaneously nonfilament and nonvolatile. It further provides additional evidence for the role of phase separation in enabling nonvolatile memory.

## *Results*

### Switching and retention of nonfilamentary $TaO_x$-based resistive memory

We fabricated $Ta_2O_5$ (15 nm)/$TaO_x$ (30 nm) cross-point resistive switching devices ranging from 20 to 60 μm laterally using optical lithography and thin-film sputter deposition. A top-view secondary electron microscopy (SEM) image of a 50 μm cross-point device is shown in Figure 1A. The detailed fabrication process is described in the Methods and Figure S1. Cross-section scanning transmission electron microscopy (Figure 1A) confirms the formation of a continuous 15-nm $Ta_2O_5$ switching layer and a 30-nm $TaO_x$ oxygen vacancy reservoir layer. Chemical composition analysis, performed using energy-dispersive X-ray spectroscopy (EDS) mapping, reveals the presence of two distinct layers with different film compositions (Figure S2).

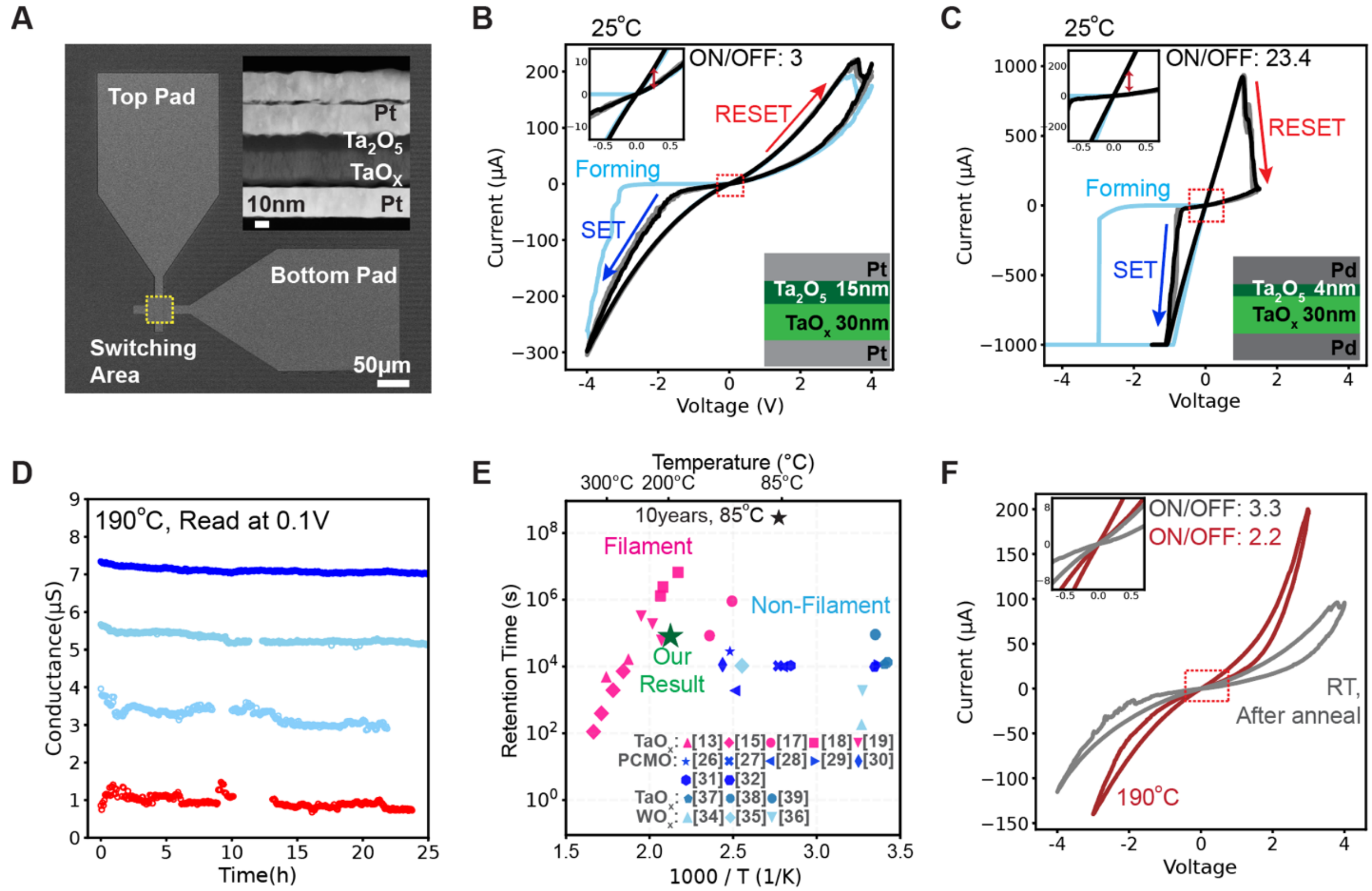

**Figure 1**. Structure and electrical characteristics of the $TaO_x$-based resistive memory device. (A)Top-view SEM image of a $TaO_x$-based resistive memory cell with a switching layer area of 50 × 50 µm². The inset cross-sectional TEM image shows the vertical stack structure of the device. (B-C) DC sweep I–V switching characteristics for this device (B) and a filamentary device (C), respectively. The inset image shows vertical stack schematics of each device. The device with a 15 nm $Ta_2O_5$ layer (B) exhibits gradual switching, whereas the one with a 4 nm $Ta_2O_5$ layer (C) shows abrupt switching. (D) Information retention test at various conductance states at 190 °C shows over 22 hours of retention. Some data loss occurred when the probe was temporarily disconnected due to laboratory vibrations. (E) Arrhenius plots of the retention time in our devices benchmarked against previous filamentary and nonfilamentary devices, including: $TaO_x$-based filamentary device (pink, solid)[13,15,17–19], PCMO-based non-filamentary (blue, open)[26–32], $TaO_x$-based non-filamentary (skyblue, solid)[37–39], $WO_x$-based non-filamentary (darkblue, solid)[34–36] industry target of 10 years at 85 °C, and our result (green, star). This graph includes only the measured results from previous papers, not extrapolated results. (F) Switching curve for a device at 190°C(brown) and RT (grey) after a 22-hour retention test at 190°C.

The pristine device shows a resistance of 1.0 GΩ when measured at 0.1V. Figure 1B shows the DC resistive switching characteristics from -4V to +4V. The first cycle (blue curve) has a slightly different profile from the subsequent cycles (black and gray curves), which show more reproducible voltage profiles. In all switching cycles, the current-voltage profile exhibits a gradual switching behavior, similar to that of nonfilamentary interfacial memory [38,39,54,55]. The resistance of the LRS is 34.6 kΩ when measured at 0.1V, while the resistance of the HRS is 102 kΩ; no compliance current was used for either SET or RESET. For comparison, Figure 1C presents an example of filamentary switching characteristics of a $Ta_2O_5$ (4 nm)/$TaO_x$ (30 nm) device measured with a 1 mA compliance[13]. The filamentary device shows a sudden change in the current during SET and RESET at the lower bias due to the formation and dissolution of a conductive filament; in contrast, our devices with a 15 nm $Ta_2O_5$ switching layer exhibits significantly smoother switching transitions (Figure 1B) that develop gradually, a result consistent with conduction modulation across a larger area as opposed to a single filament. We will present additional evidence of nonfilamentary switching later in the text.

Having shown a gradual, non-abrupt switching curve, we next consider its ability to retain information over time. We conducted retention measurements at a 0.1V read bias and under vacuum at 190 °C for 22 hours. We first programmed four devices to different conductance levels: 1 μS, 3.5 μS, 5.5 μS, and 7.2 μS. During annealing, the conductance of each device remains well separated (Figure 1D). After 22 hours at 190 °C, the programmed conductance states changed by less than 10%, showing stable retention under elevated-temperature conditions. Moreover, rather than converging, the conductance remains essentially unchanged over the 22 hours; this lack of retention failure means that the activation energy cannot be computed.

Figure 1E compares the retention of our result with previously reported filamentary and nonfilamentary two-terminal valence-change memory. While not exhaustive, the literature reveals a consistent trend: filamentary devices with abrupt switching often exhibit excellent retention, whereas nonfilamentary devices with gradual switching display poor retention. The retention performance of our device is comparable to that of filamentary devices[13,15,17–19] and significantly exceeds that of nonfilamentary devices; no other previous nonfilamentary two-terminal device has demonstrated retention that matches the time-temperature Pareto of the filamentary devices. Such filamentary devices have extrapolated 1 day at 190 °C to well over 10 years at 85 °C. Although we did not measure our devices for 10 years, we believe they can meet a similar benchmark based on the high-temperature performance.

Our devices are still operational after the 22-hour retention test at 190°C. Figure 1F presents switching data for a device after the retention test; this device was switched at both room temperature and at 190°C. This result further indicates that the device can both switch and retain information over a temperature range of room temperature to 190 °C, which makes it useful for extreme environment applications.

**Composition depth profile display bulk switching**

The previous section presented a nonvolatile ReRAM device that appears to be nonfilamentary based on its current-voltage behavior. Here, we further investigate whether this device is nonfilamentary based on 3-D depth profiling using time-of-flight secondary ion mass spectrometry (ToF-SIMS) depth profiling of the negative O- ions. The depth profiles for the other elements for the HRS devices are shown in Figure S4.

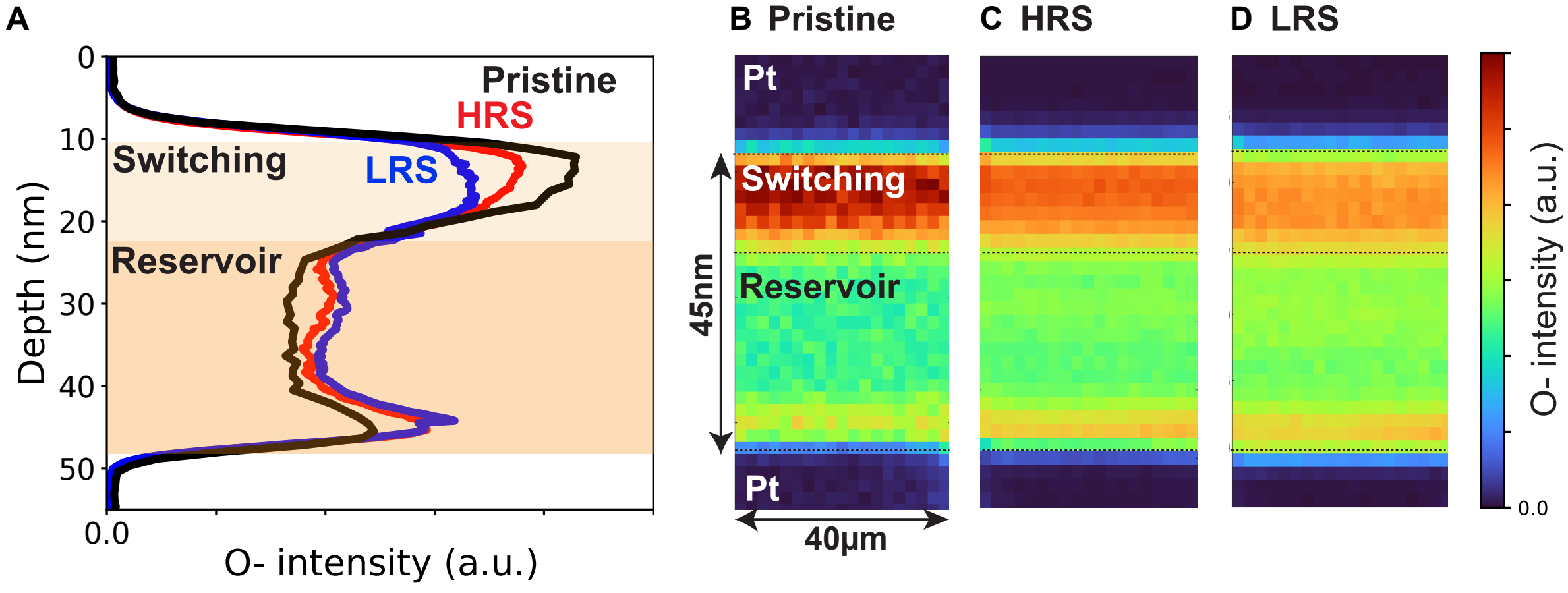

**Figure 2.** ToF-SIMS depth profile of our $TaO_x$-based resistive memory devices in pristine, low resistance, and high resistance states. (A) One-dimensional depth profiles of $O^-$ ion signals for pristine (black), HRS (red), and LRS (blue) samples averaged over the entire device area. (B-D) The corresponding two-dimensional $O^-$ ion maps of the pristine, HRS, and LRS devices show spatially uniform changes in the oxygen concentration consistent with changes in the electronic resistance. For improved signal-to-noise, these 2D images represent the average O- intensity of 75 slices in the x-z direction; representative single-slice data are shown in Figure S5.

Figure 2A shows the average depth profiles of 40 × 40 μm² devices in the pristine, LRS, and HRS states. The pristine device (black, ~1 GΩ) exhibits the highest oxygen signal within the $Ta_2O_5$ switching region and the lowest in the $TaO_x$ reservoir. When programmed to the LRS (blue, 56.3 kΩ, 17.8 μS; I-V curve in Figure S3A), the oxygen intensity decreases in the switching layer and increases in the reservoir layer. This indicates that oxygen migrated from the upper $Ta_2O_5$ layer into the lower $TaO_x$ layer during switching, consistent with a decrease in resistance as the switching layer is reduced. The HRS state (red, 162.3 kΩ, 6.2 μS; I-V curve in Figure S3B) exhibits an increase in the oxygen ion signal in the switching layer and a slightly more reduced $TaO_x$ reservoir compared to the LRS, consistent with a final resistance between the LRS and the pristine. These results confirm that the average oxygen concentration in the bulk changes during switching.

To confirm the lateral distribution of oxygen migration, we present 2D depth-resolved colormaps generated by projecting the 3D datasets onto the X–Z plane (Figures 2B–2D). All colormaps share a consistent color scale to represent the oxygen anion concentration. Consistent with the 1D averaged depth profiles, the pristine device (Figure 2B) exhibits the highest oxygen concentration within the switching layer. The LRS (Figure 2D) exhibits the lowest oxygen signal in the $TaO_x$ region, while the HRS (Figure 2C) lies between the pristine and LRS states. Importantly, these lateral maps again reveal that oxygen migration occurs throughout the full lateral dimensions of the switching layer with no evidence of filament formation. Figure S5 shows some slices extracted from the full 3D dataset, which further confirm the bulk switching and the absence of filament formation, despite the poor signal-to-noise from single-pixel measurements.

The ToF-SIMS depth profiles provide additional evidence that switching is not controlled by a single dominant filament, as in filamentary ReRAM. No filaments were detected among any of the slices within the 3D datasets; however, this measurement cannot rule out the possibility of filaments smaller than the ToF-SIMS lateral pixel size of approximately 1 μm. Instead, the more compelling evidence for nonfilamentary switching is the changes in the bulk oxygen ratios between the $TaO_x$ reservoir and initially $Ta_2O_5$ switching layer in both the 1D depth profile (Figure 2A) and the 2D projected maps (Figures 2B-D). A single nanosized filament on the order of 10 nm[13] would not provide enough signal to change the average profile of a 40-μm device that is over $10^7$ times larger. Instead, the ToF-SIMS depth profiles indicate that the bulk of the switching layer becomes more oxidized and reduced during switching in a manner consistent with the change in the device resistance. Given that the I-V profiles show non-Ohmic behavior for all states (Figure 1B), we propose that the changes in the bulk oxygen concentration change the Schottky barrier width to allow for greater electron tunneling at the Pt/$TaO_x$ interface[56].

## Area-Dependent Switching Behavior

To further assess whether the device is nonfilamentary, we investigate how its conductance depends on device area. If conduction is dominated by a single nanosized filament, the current should be independent of the device's size, while the apparent current density decreases with smaller devices. In contrast, if conduction is nonfilamentary, then larger devices should have a higher current, while the current density is independent of size[39]. To investigate this behavior, we fabricated and switched devices of different sizes, ranging from 20 × 20 $\mu m^2$ to 60 × 60 $\mu m^2$. Figures 3A–3E present SEM images of the different device sizes along with their corresponding switching curves. Regardless of the device size, all devices exhibit smooth and reproducible resistive switching behavior across the cycles ($2^{nd}$ – $4^{th}$), without the abrupt changes in current associated with filamentary devices (Figure 1C). Additionally, smaller devices show substantially less current than larger ones: in the LRS, the 20 × 20 $\mu m^2$ show 0.091 µA at 0.1V and the largest 60 × 60 $\mu m^2$ show 3.35 µA; for the HRS, the values are 0.017 µA and 0.582 µA, respectively. In Figure 3F, we plot the current measured at 0.2V as a function of the switching layer size. With 5-6 devices per size, our results show that the average current for both the HRS and the LRS is proportional to the device area ($R^2 = 0.98$). This result is expected in area-dependent devices, but not in filamentary devices where conduction is dominated by a single filament. Consistent with the shape of the voltage sweeps and the compositional depth profiles (Figure 2), this area-dependent current indicates that the current does not flow through filaments but instead spreads across the entire switching area, in agreement with a uniform, bulk-dominated conduction.

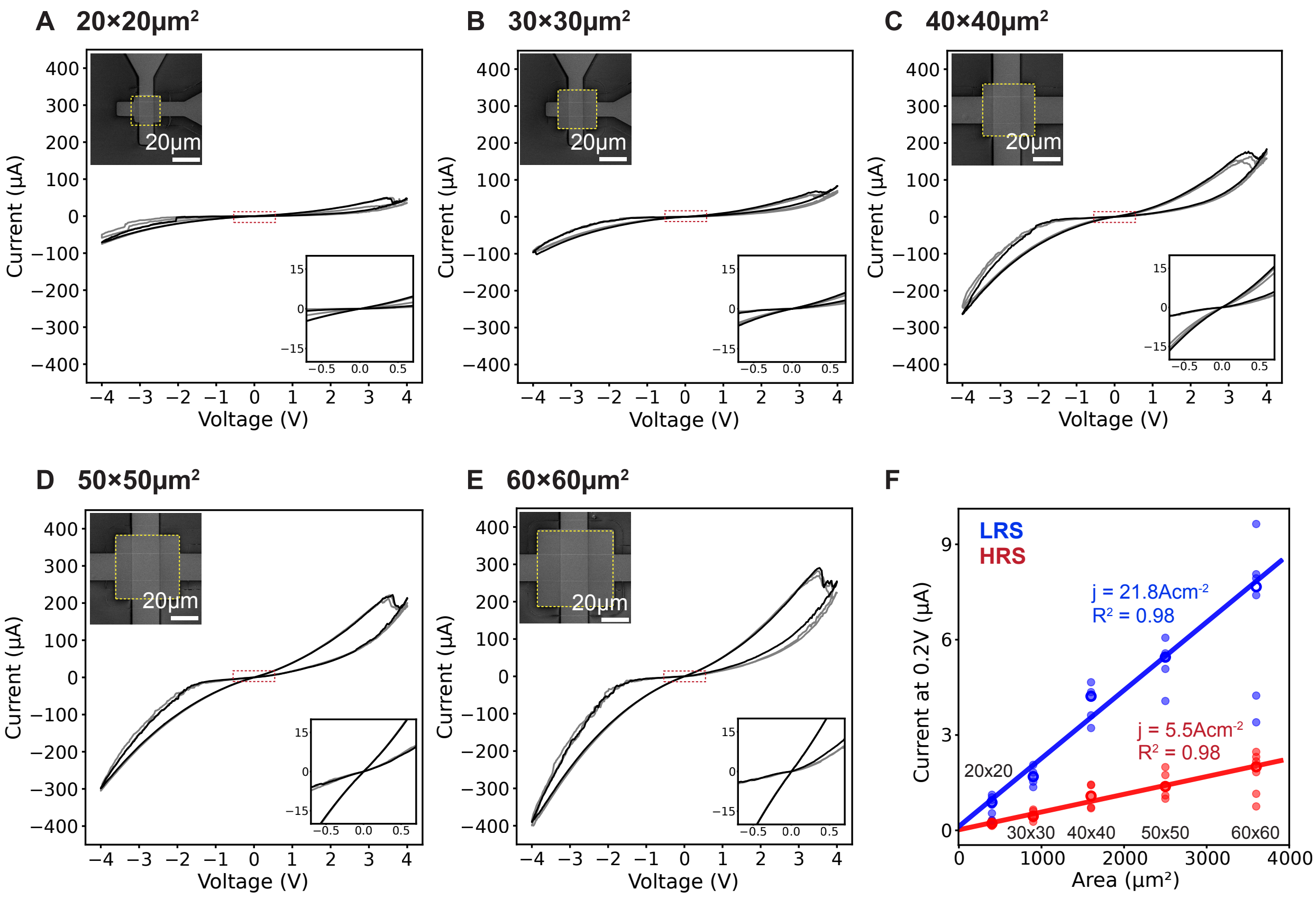

**Figure 3.** Effect of device area on the device currents. (A-E) DC sweep I–V switching characteristics for each device size, where each dataset represents three consecutive switching cycles. The inset SEM images show the corresponding device dimensions; all images and axes are displayed at the same scale. (F) The current of the high-resistance state (HRS) and low-resistance state (LRS) is proportional to the device area, signifying a nonfilamentary, area-dependent device.

## Multi-level switching and device endurance

To further evaluate the analog switching characteristics, we investigated the conductance modulation under similar incremental step-pulse programming. In this scheme, the SET programming pulse (500μs) is gradually increased from 1.4V to 3.8V. A long -4.0V RESET erase pulse (500 ms) is applied after each SET programming pulse. After every SET/RESET pulse, the program status is read at 0.1V (Figure 4A). This method allows control of the device conductance and enables multi-level operation.

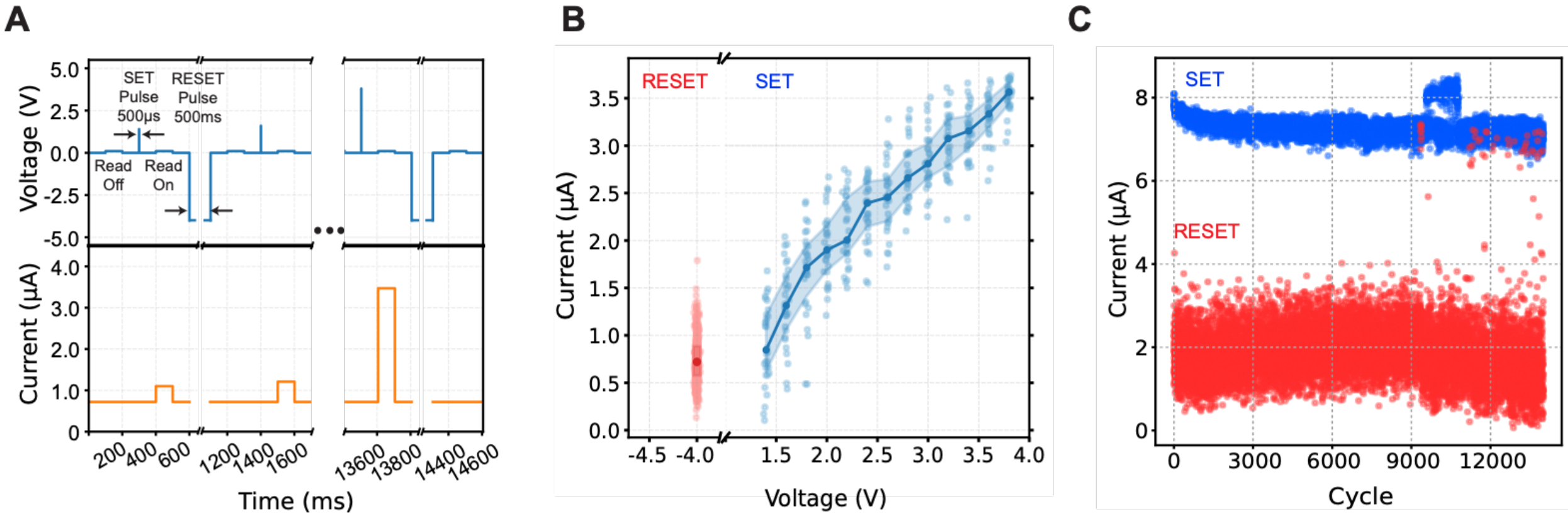

**Figure 4.** Multi-level switching and endurance of the nonfilamentary resistive memory device**.** (A) SET/RESET Incremental step-pulse programming scheme. The SET pulse voltage is increased from 1.4 to 3.8 V with a 0.2 V step (500 µs pulse width), followed by a -4.0V RESET pulse (500 ms). The device state is read at 0.1 V after each SET/RESET operation. (B) Current response at 0.1V as a function of SET pulse voltage measured over 30 repeated cycles, showing gradual conductance modulation and continuous analog switching behavior. (C) Endurance characteristics result for the first 14,000 cycles. SET was conducted using -3.5 V for 500 µs; RESET was conducted 4.0 V for 10 ms.

Figure 4B shows multi-level switching results, the current response as a function of the SET pulse voltage. To conduct this test, the switching scheme in Figure 4A pulse scheme was repeated 30 times for 390 total SET/RESET pairs. As the SET pulse voltage increases, the device conductance increases progressively, confirming gradual analog switching behavior. The cycles exhibit smooth and continuous current evolution without abrupt current jumps.

Our results reveal two limitations. First, the programming and erase speeds are relatively slow. The minimum program speed of our device is 10µs, and the minimum erase speed is 100µs (Figure S6). To obtain stable erase and program states, we use a pulse time longer than the limit; this speed is comparable to that of a $TiO_x$ nonfilamentary device by Zhou et al[47]. This slow program behavior originates from slow oxygen ion migration within the switching layer; due to the absence of a filament, no Joule heating to accelerate switching is expected[57]. Second, the programmed conductance levels partially overlap with neighboring analog states, reducing the accuracy of intermediate states. Iterative write/verify algorithms could be used to improve state precision[58].

Finally, to examine device reliability, we perform device endurance measurements for the first 14,000 SET/RESET cycles (Figure 4C). During the initial ~2,000 cycles, the SET current decreases by approximately 7%, after which it stabilizes and remains nearly constant throughout the remaining cycles. In the first 9,000 cycles, there are no switching failures. Afterwards, however, some of the RESET failed to switch. The high switching voltages may be responsible for the relatively low device endurance.

## *Discussion*

The stable retention and area-dependent switching demonstrated by our $Ta_2O_5$-based resistive memory provide strong evidence that the device overcomes the conventional trade-off between separated filamentary and non-filamentary memristors. Filamentary memristors are believed to show long retention thanks to deep, energetically stable conductive filaments, but at the cost of high currents, large device-to-device variability, and stochastic switching. Non-filamentary memristors, by contrast, achieve smoother modulation and size-dependent switching but lose their programmed states under elevated temperatures and long retention time. We successfully demonstrated a device that achieves the high thermal retention of filamentary memristors while simultaneously exhibiting the spatially uniform conduction characteristics associated with non-filamentary memristors. The robust multi-level retention observed at 190 °C for more than 22 hours constitutes a significant improvement over earlier non-filamentary platforms and strongly implies that the switching mechanism is not restricted to a narrow Schottky interface but instead involves a more spatially extended process.

We propose that the origin of this long retention lies in the phase separation of tantalum oxide. The composition depth profiles (Figure 2) show two regions of different oxygen concentration and therefore different metal-to-oxygen ratios. However, because tantalum oxide undergoes phase separation, oxygen would not diffuse from high to low concentrations. Instead, the simultaneous existence of two compositions is the thermodynamically stable configuration[13]. This phase separation of amorphous tantalum oxide has been definitively demonstrated in our previous work on two-terminal filamentary resistive memory[13] and three-terminal electrochemical memory [49]. We propose that this phase separation also enables nonvolatile information retention in nonfilamentary, two-terminal memory devices. This design rule of utilizing phase separation may also apply to other interfacial memristors with poor information retention, such as those using 2D materials[59]. By showing that even interfacial memory can be nonvolatile, our results support the strategy of using phase separation to improve the retention of bulk electrochemical memory with faster-diffusing ions, such as those using lithium[60], protons[61], and copper[62].

Our work shows that there is no fundamental tradeoff between nonfilamentary switching and long retention time, as previously believed. However, substantial amounts of engineering are necessary for improvement. The speed, multilevel switching, and endurance characteristics are not as good as state-of-the-art devices. While our devices exhibit area-dependent behavior, we have not been able to fabricate devices smaller than 20 μm. Our devices are also strongly dependent on processing conditions. For example, while using 200W of sputter power showed excellent devices, devices using 50 or 100 W show either no switching or filamentary switching (Figure S7).

## *Conclusion*

In summary, we develop a $Ta_2O_5/TaO_x$ resistive memory that is simultaneously nonfilamentary and nonvolatile, thereby solving one of the longstanding challenges in reistive memory. This device shows smooth switching, area-dependent current scaling, and long-term stability even under high temperatures. The comprehensive electrical analysis, area-dependency measurements, and ToF-SIMS chemical profiling all converge to provide a coherent picture of a bulk-type switching regime. This mechanism enables deterministic modulation of conductance across the entire oxide layer, yielding robust multi-level states and strong retention performance. The insights gained from this study establish a new materials and device design strategy for high-uniformity, high-stability ReRAM, offering a promising platform for next-generation neuromorphic and non-volatile memory applications.

## *Experimental Procedures*

### Fabrication of $TaO_x/Ta_2O_5$ Cells

Four-inch silicon wafers with a 500-nm-thick thermally grown $SiO_2$ layer (UniversityWafer) were used as substrates. A bilayer photoresist stack was sequentially spin-coated at 4000 rpm for 30 s, consisting of LOR 10B (≈10 μm) and SPR 955 (≈800 nm). The bottom electrode pattern was defined using a Heidelberg μPG 501 mask writer. Subsequently, a 30-nm-thick platinum (Pt) layer was deposited by DC sputtering in an AJA Orion-8 system using a 2-inch Pt target (Plasmaterials, Livermore, CA). The deposition was performed at a target power of 100 W under 3 mTorr of pure argon, with a target-to-substrate distance of approximately 15 cm. All sputtering was conducted at room temperature without any post-deposition annealing. The lift-off process was carried out by immersing the patterned samples in Remover PG for more than 8 h, followed by rinsing with isopropyl alcohol (IPA) and deionized (DI) water.

The switching layer patterns, ranging from $10 \times 10$ μm² to $100 \times 100$ μm², were defined using the same lithographic procedure. A multilayer stack consisting of a 45-nm-thick substoichiometric $TaO_x$ layer, a 15-nm-thick stoichiometric $Ta_2O_5$ layer, and a 30-nm-thick Pt passivation layer was deposited sequentially in the same sputtering system without breaking vacuum. The target power was set to 100 W, 200 W, and 100 W for the $TaO_x$, $Ta_2O_5$, and Pt layers, respectively. The substoichiometric $TaO_x$ layer was deposited under an Ar:$O_2$ gas mixture (85:15 ratio), while the stoichiometric $Ta_2O_5$ layer was deposited under a 50:50 Ar:$O_2$ mixture, both at a total pressure of 3 mTorr controlled by mass flow controllers. The lift-off process was then performed under identical conditions as described above. Finally, the top electrode patterns were defined, and a 30-nm-thick Pt was deposited following the same sputtering and lift-off procedures.

**Device Measurements**

Electrical measurements of the $Ta_2O_5$ interfacial memory devices were performed using a Keithley 4200 semiconductor parameter analyzer. For DC switching characterization, a voltage bias was applied to the top electrode while the bottom electrode was grounded. The current compliance was limited to 1 mA to prevent permanent dielectric breakdown. Each device was subjected to five consecutive switching cycles, consisting of SET and RESET operations with voltage sweeps of −4 V and +4 V, respectively. The initial switching event was always in the SET direction. Representative DC I–V switching characteristics are shown in Figures 1B and 3.

Retention measurements (Figures 1D and 1E) were conducted after 20 pulse-cycling operations. Annealing was performed in a temperature- and environment-controlled probe station (Everbeing CG-196) at 190 °C under a vacuum of approximately 40 mTorr, following two argon purge cycles. During retention testing, each state was switched at 4 V (RESET), -2 V (LRS 1), -3 V(LRS 2), and -4 V (LRS 3). During the test, the channel conductance was recorded every 3 minutes for over 22 hours using the Keithley 4200 system.

The multilevel switching behavior (Figure 4B) of the $Ta_2O_5$ device was evaluated using the Keithley 4200 PMU system. Endurance testing (Figure 4D) was carried out for approximately 14,000 switching cycles using voltage pulses generated by the Keithley 4200. The SET and RESET pulse conditions were −3.5 V for 500 µs and +4.0 V for 10 ms, respectively.

## Material Characterization (TEM and SEM)

Three samples—low-resistance state (LRS), high-resistance state (HRS), and pristine—were prepared for scanning transmission electron microscopy (STEM) analysis using a Thermo Fisher Helios 650 Ga+ ion plasma focused ion beam (FIB) system at the University of Michigan. Samples were extracted and thinned using 30 kV $Ga^+$ ions, followed by final polishing at 2 kV, 0.19 nA to minimize surface damage. STEM imaging and energy-dispersive X-ray spectroscopy (EDS) analyses were performed using a Thermo Fisher Talos F200X G2 microscope operated at 200 kV in STEM mode. All image and spectral data were processed and analyzed using *Velox* software.

## ToF SIMS Measurements

The time-of-flight secondary ion mass spectrometry (ToF-SIMS) analysis was performed using the ToF.SIMS.5-NSC instrument (ION.TOF GmbH) at the Center for Nanophase Materials Sciences at Oak Ridge National Laboratory. A $Bi^{3+}$liquid metal ion gun, operating at 30 keV energy, 0.5 nA current (DC mode), and with a spot size of approximately 120 nm, served as the primary source for chemical analysis. A $Cs^+$ sputter ion gun was additionally used with operating at 1 keV energy and 70 nA current for depth profiling. The measurements were conducted in non-interlaced mode, with each analysis scan by $Bi^{3+}$ (100 × 100 $\mu m^2$) being succeeded by 2 seconds of sputtering with $Cs^+$ (300 × 300 $\mu m^2$). A low-energy electron flood gun was used for charge compensation. Secondary ions were then analyzed using time-of-flight mass analyzers with a mass resolution of $m/\Delta m$ = 100–300 in the negative ion detection mode. Intensities of the peaks corresponding to $O^-$. Data normalization was conducted to compare the data between different states; each intensity data point of $O^-$ was divided by the integrated area intensity of some of the switching and reservoir layers. To improve the signal-to-noise, the data were binned such that the pixel sizes are 1 μm × 1 μm × 2 nm.

## Resource Availability

***Lead Contact:*** Requests for further information and resources should be directed to and will be fulfilled by the lead contact, Yiyang Li, yiyangli@umich.edu

***Materials Availability:*** This work did not generate new unique reagents or materials.

***Data and Code Availability:*** If this article is accepted, the data for this work will be published on the Deep Blue Repository, with the DOI linked to this Article.

## Acknowledgement

This work was supported by a Defense Advanced Research Project Agency (DARPA), Young Faculty Award no. D24AP00328. S. L. was supported by a SK Hynix PhD scholarship. The authors acknowledge the Michigan Center for Materials Characterization for the use of the instruments and staff assistance. The fabrication of the microelectrode arrays was performed at the University of Michigan Lurie Nanofabrication Facility. ToF-SIMS characterization was conducted at the Center for Nanophase Materials Sciences, which is a DOE Office of Science User Facility, and using instrumentation within ORNL's Materials Characterization Core provided by UT-Battelle, LLC under Contract No. DE-AC05-00OR22725 with the U.S. Department of Energy.

## Author contributions

S.L. designed and fabricated the device and performed electrical testing. A.I conducted ToF-SIMS analysis. D.S. and J.L. assisted with device design and data analysis. Y.L. supervised the project. All authors reviewed and approved the manuscript.

## Declarations of Interest

The authors declare no conflicts of interest.

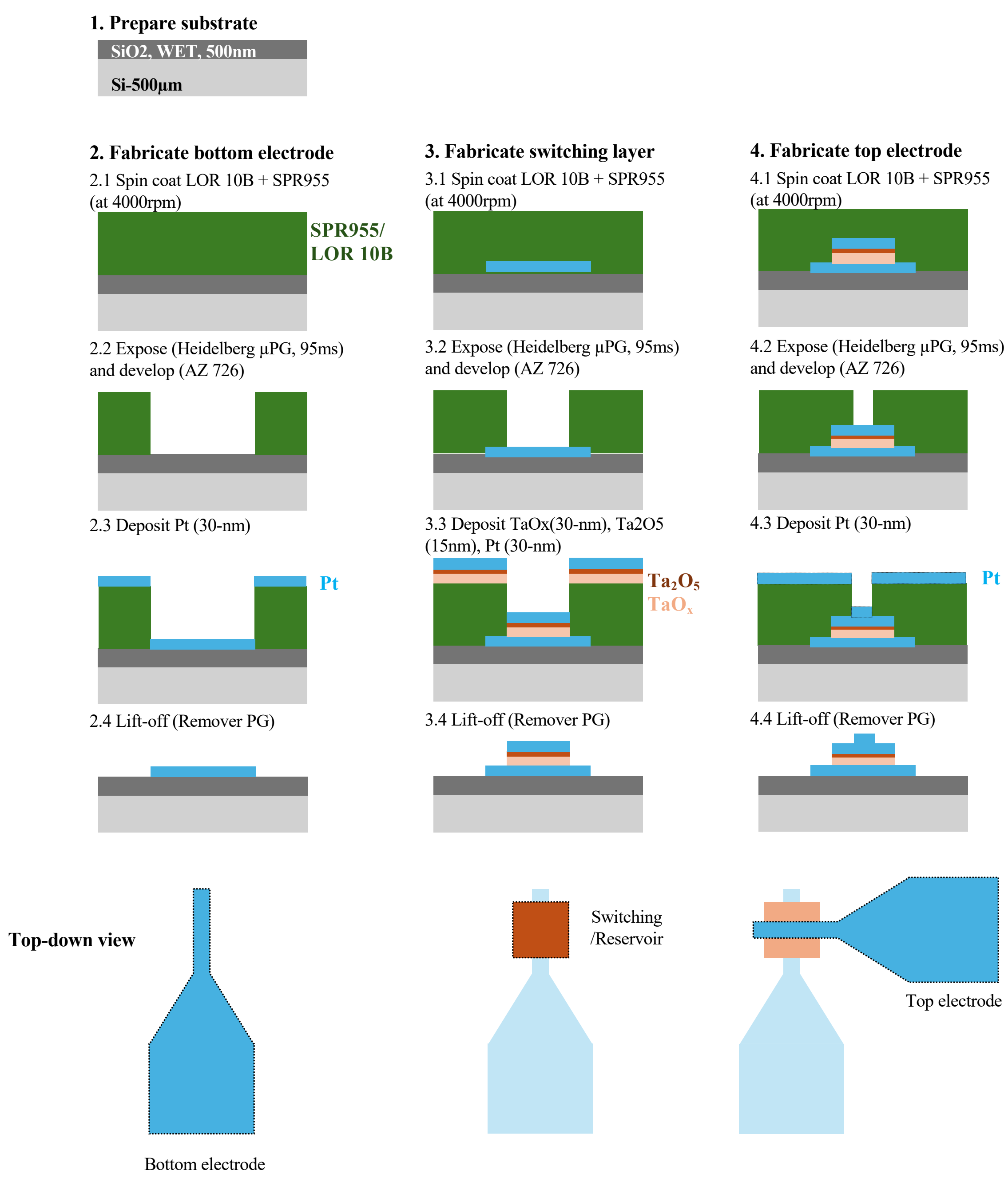

Figure S1. Process flow for the photolithography of nonfilamentary tantalum oxide devices.

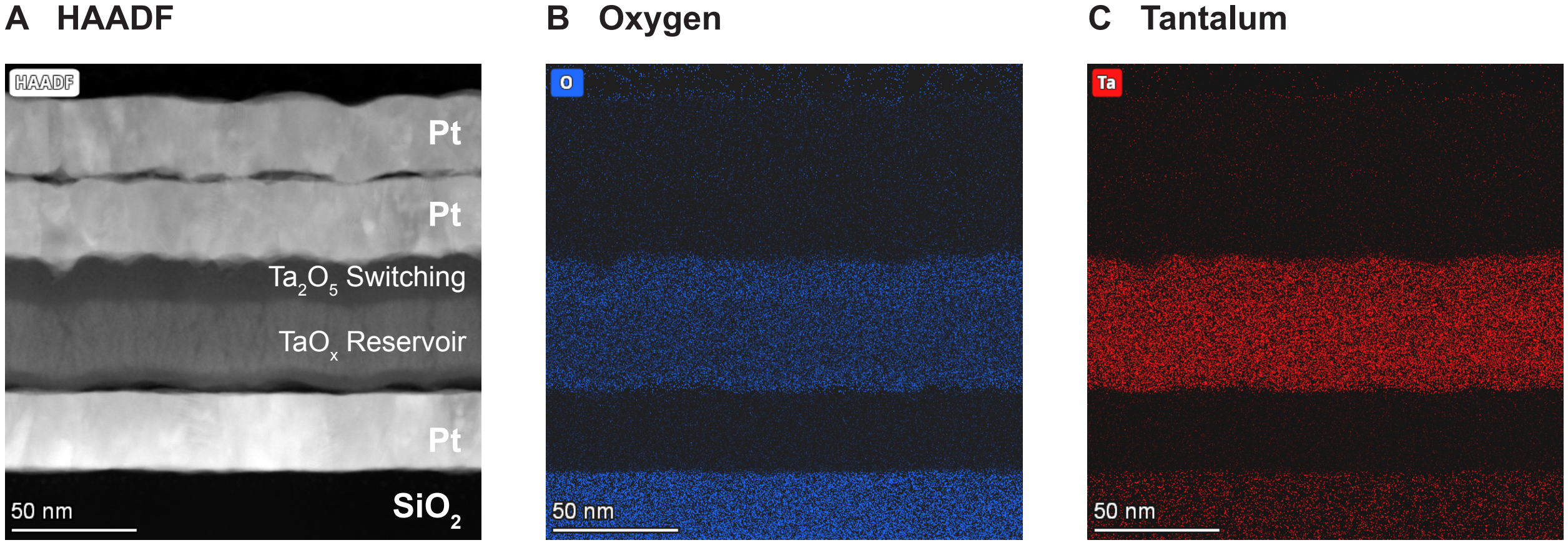

Figure S2. Cross-section scanning transmission electron microscopy of a tantalum oxide resistive memory device.

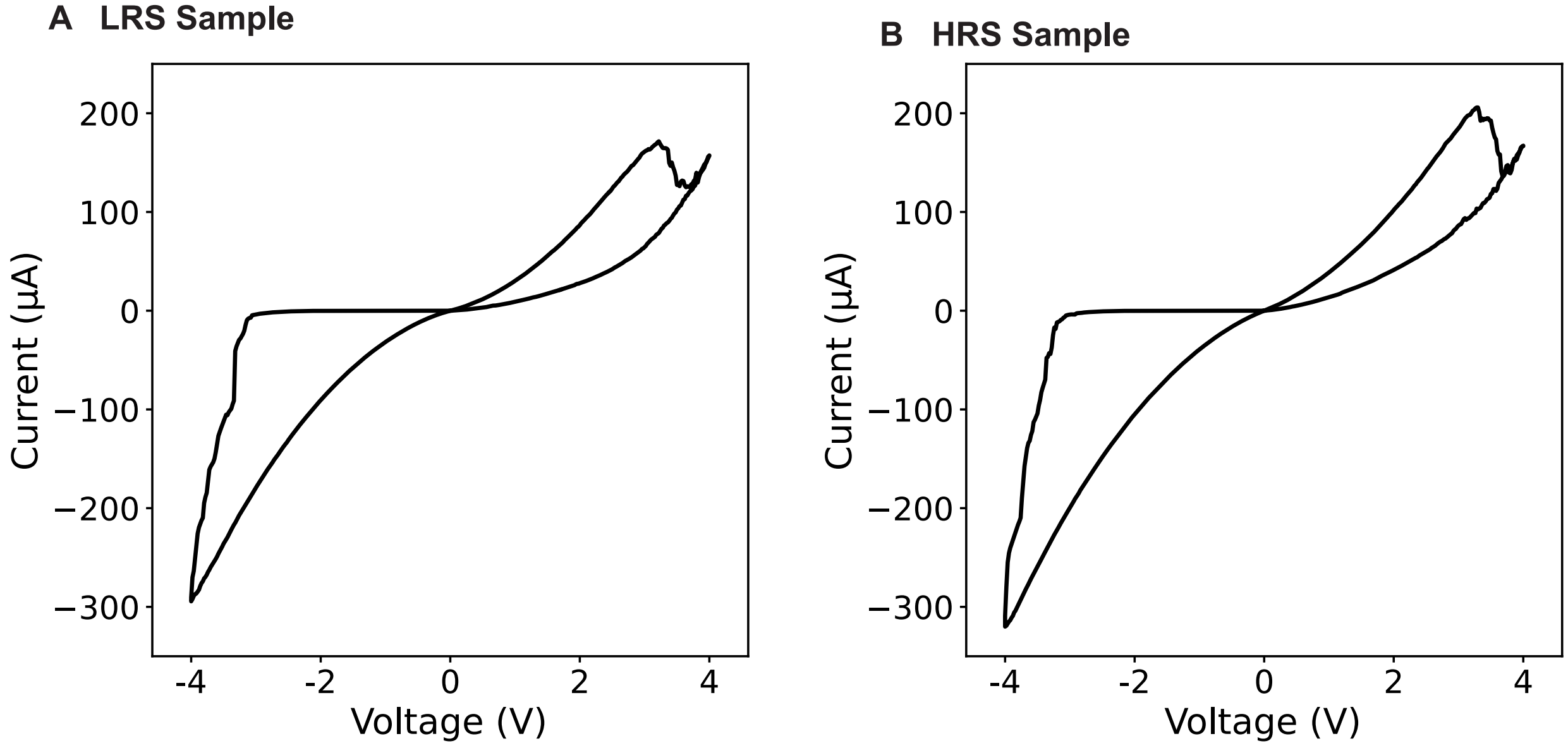

Figure S3. Current-voltage switching profile of the ToF-SIMS samples. Each device was $40 \times 40\ \mu m^2$.

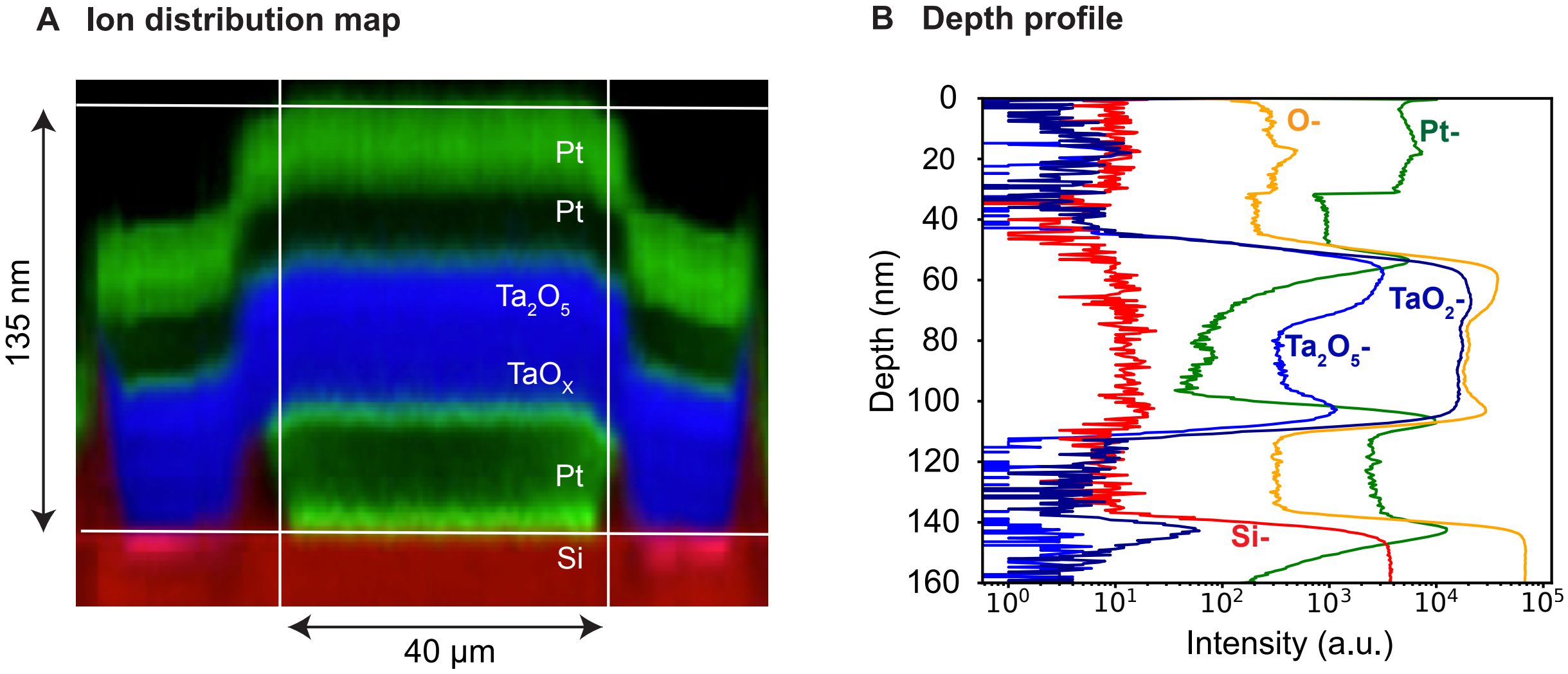

Figure S4. (A) Ion distribution map showing the spatial distribution of key species across the device cross-section. The layer structure includes Pt, $Ta_2O_5$, $TaO_x$, and Si. (B) Depth profiles of ion intensities as a function of depth; the color of each ion corresponds to that used in panel (A).

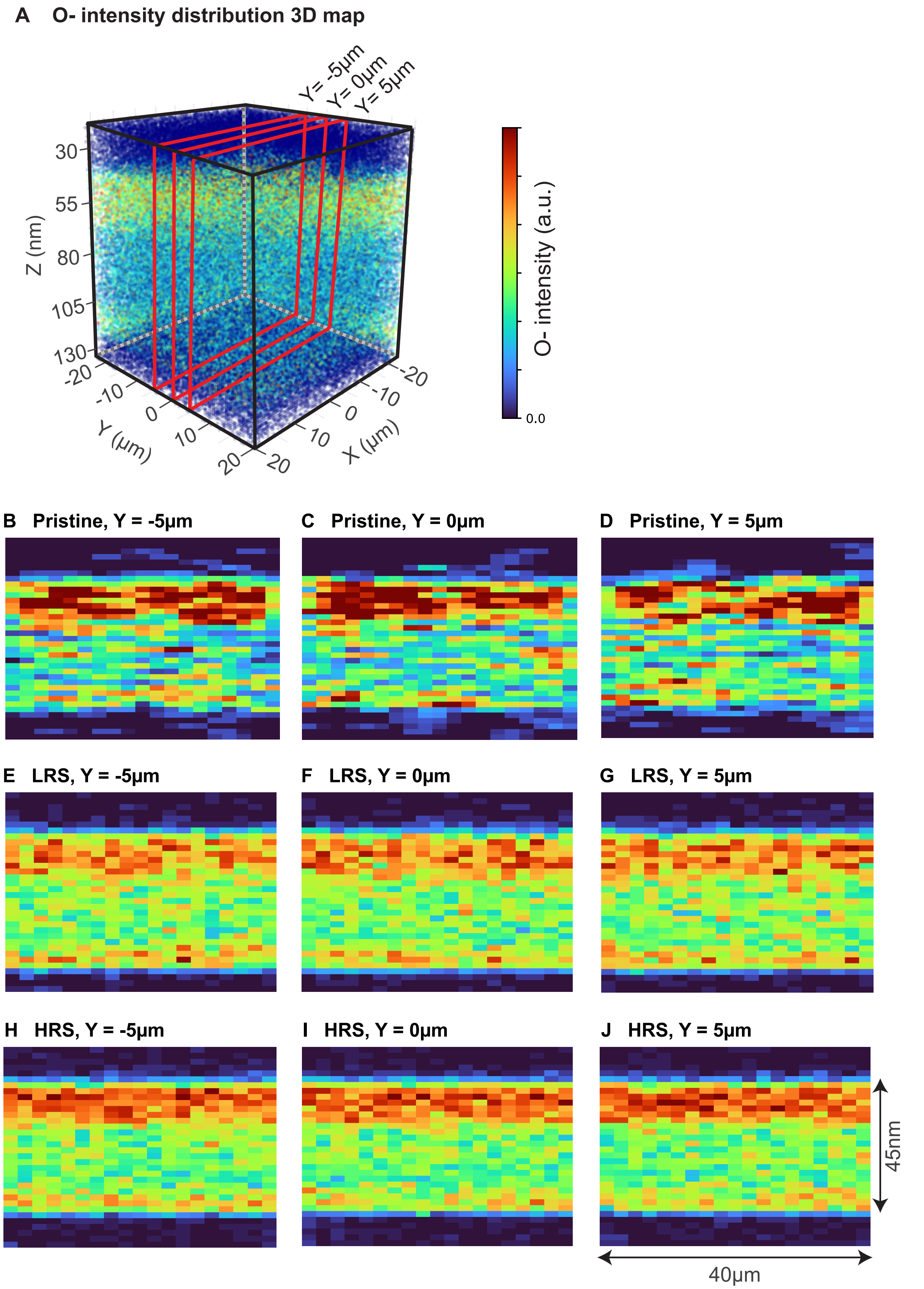

**A O- intensity distribution 3D map**

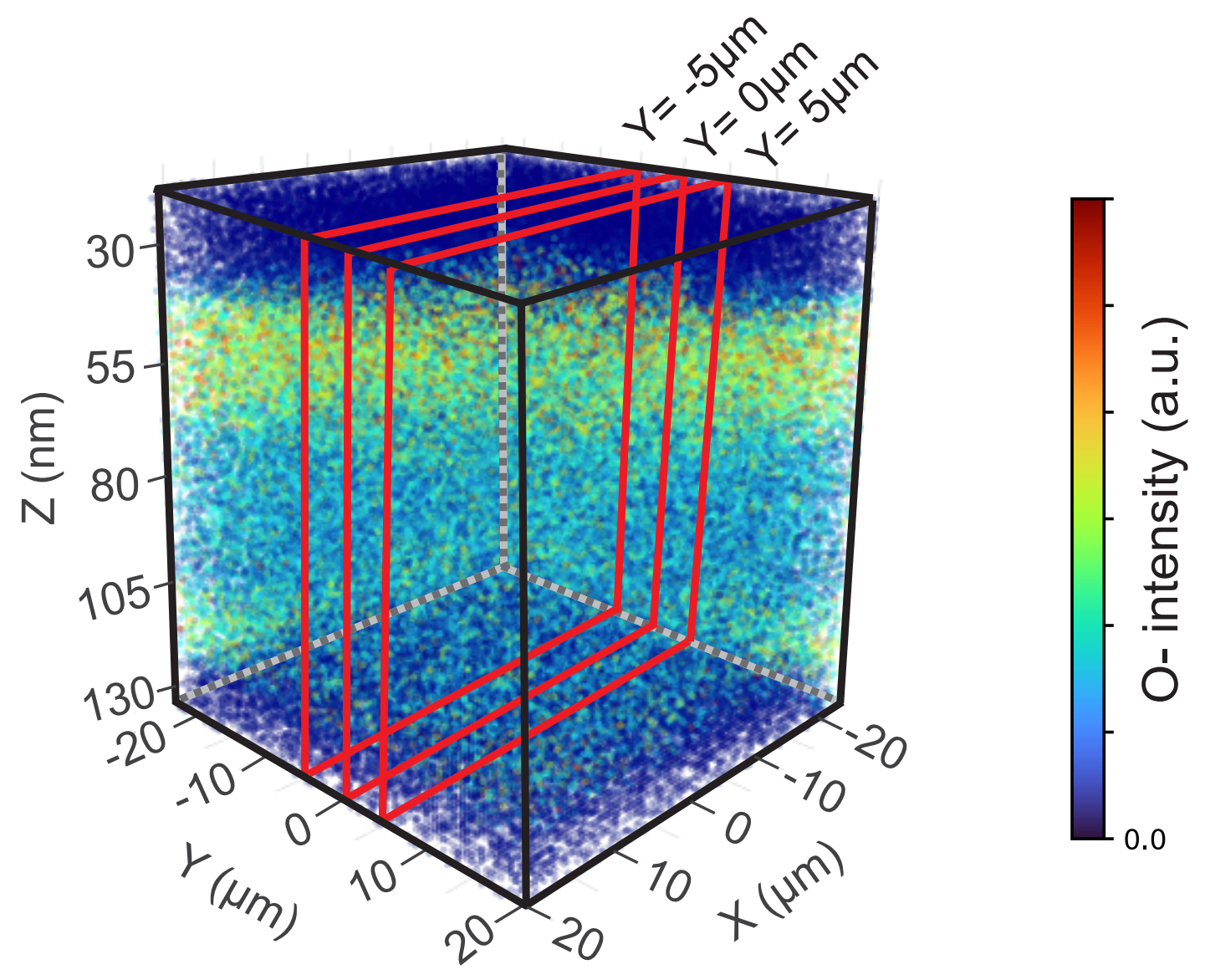

**B Pristine, Y = -5μm**

**C Pristine, Y = 0μm**

**D Pristine, Y = 5μm**

**E LRS, Y = -5μm**

**F LRS, Y = 0μm**

**G LRS, Y = 5μm**

**H HRS, Y = -5μm**

**I HRS, Y = 0μm**

**J HRS, Y = 5μm**

45nm

40μm

Figure S5. O- intensity of individual y-slices of the devices under different resistance states used to generate Figure 2. (A) Three-dimensional map of $O^-$ ion intensity, where the red boxes indicate the slicing planes corresponding to panels (B-J) at Y = -5μm, 5μm, 0μm, and 5μm. (B-D) Pristine device. (D-G) LRS device. (H-J) HRS device. All panels (B-J) used the same scale and layer structure as Figure 2.

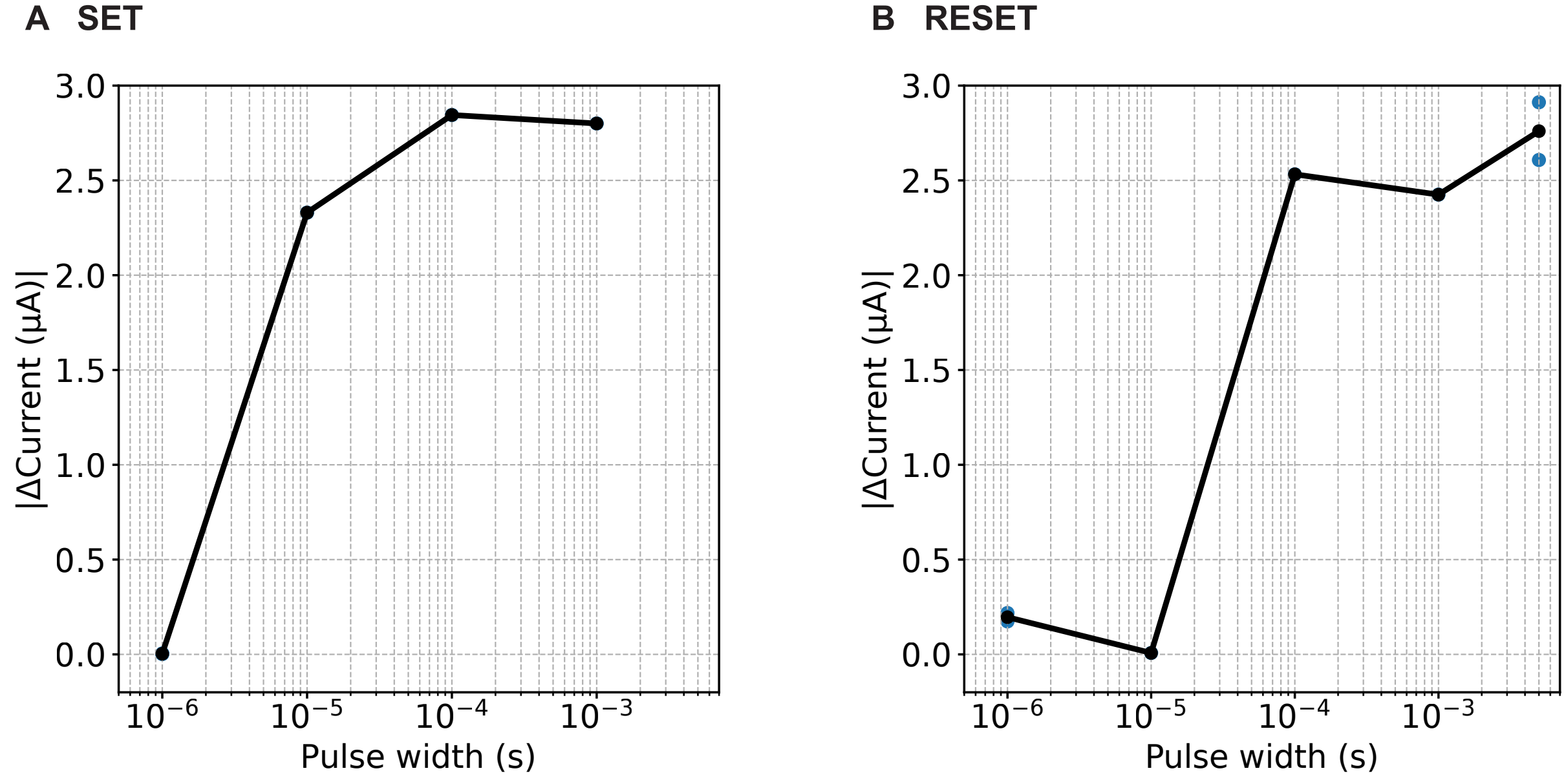

Figure S6. Programming speed of the device. The y-axis represents the absolute current change before and after the pulse, and the x-axis is the pulse width for SET and RESET operations. (A) SET switching occurs at pulse widths longer than 1μs. (B) RESET switching requires pulse widths longer than 10μs.

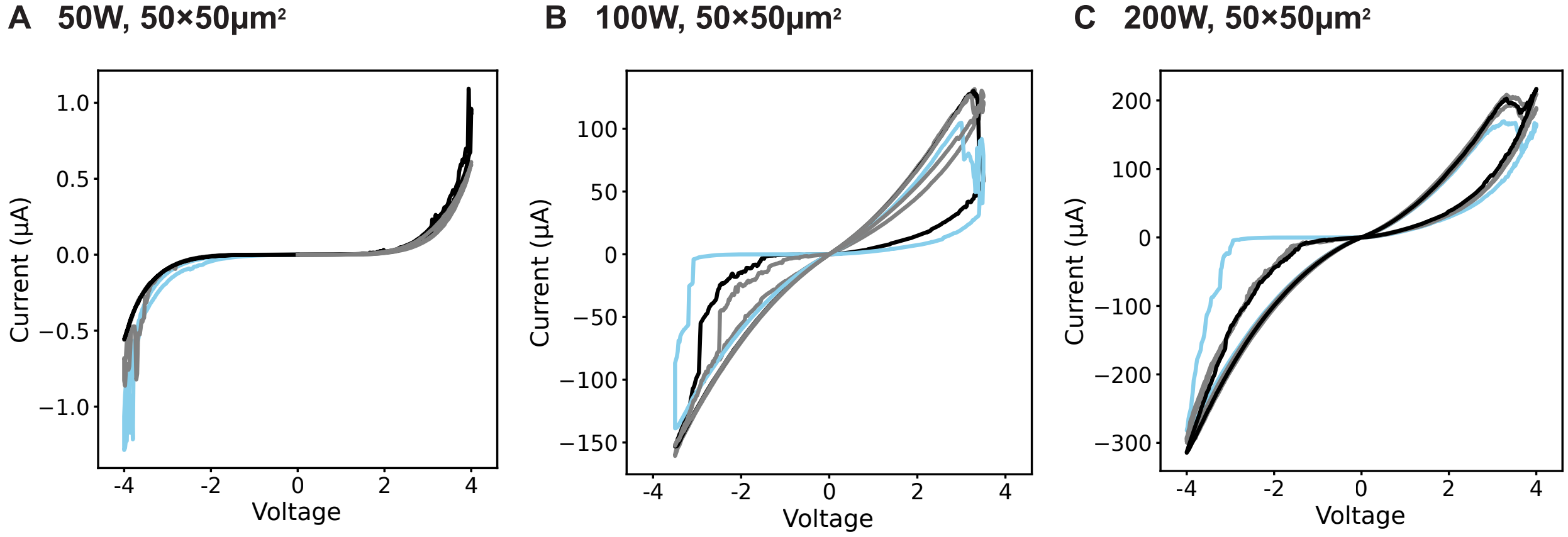

Figure S7. Switching the I-V curve of each Switching layer deposition power. (A) 50W for 4000sec, (B) 100W for 1740s, (C) 200W for 720s. The sky-blue line is the forming sweep. The black line represents the first sweep, and the grey lines represent subsequent sweeps. Only the 200W device show nonfilamentary switching behavior.